\documentclass[sigconf]{aamas} 

\usepackage{balance} 

\setcopyright{ifaamas}
\acmConference[AAMAS '26]{In 7th International Workshop on Agents for Societal Impact (ASI 2026) @ AAMAS 2026}{May 25, 2026}
{Paphos, Cyprus}{C.~Amato, L.~Dennis, V.~Mascardi, J.~Thangarajah (eds.)}
\copyrightyear{2026}
\acmYear{2026}
\acmDOI{}
\acmPrice{}
\acmISBN{}

\usepackage{booktabs}
\usepackage{graphicx}
\usepackage{subcaption}
\usepackage{tikz}
\usepackage{pgf-pie}
\usepackage{pgfplots}
\pgfplotsset{compat=1.7}
\usetikzlibrary{shapes.geometric}
\usetikzlibrary{shapes.arrows}
\usetikzlibrary{positioning}
\usepackage{amsmath}

\usepackage{multirow}
\usepackage{balance}
\usepackage{url}
\usepackage{array}

\acmSubmissionID{<<submission id>>}

\title[AAMAS-2026 Formatting Instructions]{Decision-Level Fusion for Robust Wearable Affect Recognition}

\author{Lokesh Singh}
\affiliation{
  \institution{University of Southampton}
  \city{Southampton}
  \country{United Kingdom}}
\email{lb5e23@soton.ac.uk}

\author{Athina Georgara}
\affiliation{
  \institution{University of Southampton}
  \city{Southampton}
  \country{United Kingdom}}
\email{ag1g24@soton.ac.uk}

\author{Jayati Deshmukh}
\affiliation{
  \institution{University of Southampton}
  \city{Southampton}
  \country{United Kingdom}}
\email{jd2r24@soton.ac.uk}

\author{Tan Viet Tuyen Nguyen}
\affiliation{
  \institution{University of Southampton}
  \city{Southampton}
  \country{United Kingdom}}
\email{tvtn1c23@soton.ac.uk}

\author{Sarvapali D. Ramchurn}
\affiliation{
  \institution{University of Southampton}
  \city{Southampton}
  \country{United Kingdom}}
\email{sdr1@soton.ac.uk}

\begin{abstract}

Automatic recognition of affective state from wearable physiology has clear societal impact for public health, preventive care, and stress-aware interventions, but real deployments require robustness to non-stationary dynamics, artefacts, and missing sensors. We study this problem on WESAD (baseline, stress, amusement), where common fixed-basis spectral features (e.g., FFT bandpower and Welch PSD) can oversmooth short-lived discriminative patterns. We propose a non-stationary pipeline that combines Fourier–Bessel Series Expansion (FBSE) with EWT data driven spectral segmentation to extract mode wise transient descriptors. For multimodal integration, we adopt decision-level aggregation over per-modality predictors and weight each modality by predictive uncertainty and modality reliability. On WESAD (15 subjects; ECG, EDA, BVP, EMG, ACC three classes) indicate that decision-level aggregation is $\approx84\%$ times at least as good as feature level aggregation, and $\approx48\%$ strictly better, suggesting improved robustness under heterogeneous and partially reliable sensing.

\end{abstract}

\keywords{Wearable affect recognition, physiological signals, Fourier–Bessel series expansion, decision-level aggregation, uncertainty-aware fusion.}

\newcommand{\BibTeX}{\rm B\kern-.05em{\sc i\kern-.025em b}\kern-.08em\TeX}

\begin{document}


\pagestyle{fancy}
\fancyhead{}


\maketitle 


\section{Introduction}

Automatic affective state detection (a person’s current subjective, psychological, and physiological state) from physiological signals is a key problem in AI for health, with applications such as continuous mental-health monitoring, stress-aware interventions, and adaptive human–computer interaction ~\cite{healey2005detecting,kreibig2010autonomic}. Among affective states, stress and amusement capture complementary high-arousal conditions and enable continuous assessment of emotional well-being using wearable sensing in naturalistic settings ~\cite{picard2000affective}. Recent advances in wearable technology make it possible to collect multimodal signals such as electrodermal activity (EDA), electrocardiography (ECG), electromyography (EMG), blood volume pulse (BVP), and motion (ACC), which reflect autonomic and somatic correlates of affect. Benchmark datasets such as WESAD facilitate systematic evaluation of stress and affect recognition under standard protocols, accelerating progress toward deployable affect-aware health systems ~\cite{schmidt2018introducing}. Despite this progress, accurate detection remains challenging because physiological time series are noisy, non-stationary, and subject-dependent, and affect-related dynamics often appear as transient, context-dependent changes rather than stable periodic patterns.

A major limitation of many pipelines is feature extraction. Common approaches rely on statistical summaries or Fourier-based representations computed over fixed windows (e.g., FFT bandpower or Welch PSD). These methods implicitly assume approximate local stationarity and can discard time-varying structure that is critical for distinguishing affective state ~\cite{huang1998empirical}. Time–frequency methods such as STFT and wavelets partially mitigate this limitation, but they retain trade-offs between time resolution, frequency resolution, and adaptability to signal-specific structure~\cite{wang2015imaging}. Motivated by these limitations, we focus on non-stationarity decomposition methods and, in particular, we propose the FBSE-EWT signal analysis method, which combines Fourier Bessel-series expansion with Empirical Wavelet Transform like data-driven segmentation to better preserve transient spectral structure in physiological windows~\cite{gilles2013empirical}. 

Beyond limitations in physiological signal analysis, wearable affect monitoring highly relies on multimodal sensing to leverage complementary physiological mechanisms. Combining various physiological signals (such as ECG, EMG, EDA, etc.) using machine learning has proven to improve the affective state detection~\cite{ayata2020emotion}. However, this introduces a second major challenge: the robust integration under sensor noise, signal failure, and missing modalities, common issues in real-life deployments ~\cite{gravina2017multi}. Early fusion ~\cite{chen2015feature} approaches and models that assume complete sensor availability can therefore fail under realistic conditions. To address this challenge, we adopt a modular approach in which each modality computes an independent prediction, followed by decision-level aggregation that accounts for predictive uncertainty. This design is suitable for agent-based decision support because it outputs calibrated belief states that can be used to trigger interventions conservatively under uncertainty.

In summary, we propose a three-stage pipeline for affective state detection from wearable physiology. First, we extract meaningful representations using feature extraction techniques tailored to non-stationary time series. Second, we train single-sensor predictors to classify sensor-specific features into affective state, enabling each modality to independently learn affect-discriminative patterns. Finally, we aggregate the multiple sensor classifiers to reach one final prediction. 
We do so by weighting each sensor’s contribution according to its predictive uncertainty in order to enhance robustness to sensor variability and partial failure. We evaluate the proposed framework on WESAD for three-class affect recognition (baseline / stress/ amusement). Our results suggest that non-stationary feature extraction and decision-level aggregation improve reliability in multimodal settings. The key contributions of this paper are as follows:
\begin{enumerate}
    \item A non-stationary representation for wearable physiology based on FBSE--EWT that preserves transient spectral structure via data-driven mode isolation.
    \item A modular per-modality prediction design and an uncertainty-aware, entropy-weighted decision fusion rule that remains stable under noisy or missing modalities.
    \item An empirical evaluation on WESAD demonstrating the benefit of decision-level aggregation across sensors.
\end{enumerate}

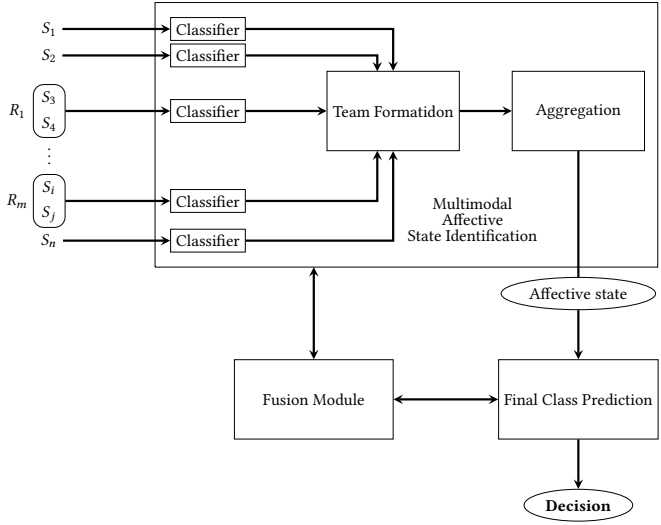
\begin{figure}[!t]\centering
\scalebox{0.7}{
    \begin{tikzpicture}
        \node (s1) at (0,4) {$S_1$};
        \node (s2) at (0,3.5) {$S_2$};
        \node (s3) at (0,2.7) {$S_3$};
        \node (s4) at (0,2.2) {$S_4$};

        \node[draw,rectangle, minimum width=6mm, minimum height = 10mm, rounded corners=0.2cm] (r1) at (0,2.45) {};
        \node at (-0.6,2.45) {$R_1$};
        
        \node  at (0,1.7) {$\vdots$};
        
        \node (si) at (0,1) {$S_i$};
        \node (sj) at (0,0.5) {$S_j$};

        \node[draw,rectangle, minimum width=6mm, minimum height = 10mm, rounded corners=0.2cm] (rm) at (0,0.75) {};
        \node at (-0.6,0.75) {$R_m$};
        
        \node (sn) at (0,0) {$S_n$};

        \node[draw,rectangle] (mlp1) at (3,4) {Classifier};
        \node[draw,rectangle] (mlp2) at (3,3.5) {Classifier};
        \node[draw,rectangle] (mlp3) at (3,2.45) {Classifier};
        \node[draw,rectangle] (mlp4) at (3,0.75) {Classifier};
        \node[draw,rectangle] (mlp5) at (3,0) {Classifier};

        \draw[-stealth,very thick](s1)--(mlp1);
        \draw[-stealth,very thick](s2)--(mlp2);
        \draw[-stealth,very thick](r1)--(mlp3);
        \draw[-stealth,very thick](rm)--(mlp4);
        \draw[-stealth,very thick](sn)--(mlp5);
        
        \node[draw,rectangle,minimum width=25mm,minimum height=15mm] (TF) at (6.5,2.45) {Team Formatidon};
        
        \draw[-stealth,very thick](mlp1)--(6.5,4)--(TF);
        \draw[-stealth,very thick](mlp2)--(6.2,3.5)--(6.2,3.2);
        \draw[-stealth,very thick](mlp3)--(TF);
        \draw[-stealth,very thick](mlp4)--(6.2,0.75)--(6.2,1.7);
        \draw[-stealth,very thick](mlp5)--(6.5,0)--(TF);
        
        \node[draw,rectangle,minimum width=25mm,minimum height=15mm] (Agg) at (10,2.45) {Aggregation};

        \draw[-stealth,very thick](TF)--(Agg);
        \node[ellipse,align=center,draw] (event) at (10,-1) {Affective state};
        \node[draw=black,ellipse, align=center] (dec) at (10,-5) {\color{black}\bf Decision};

        \node[draw=black,rectangle,minimum width=30mm,minimum height=15mm] (DM) at (10,-3) {\color{black}Final Class Prediction};
        
        \node[draw=black,rectangle,minimum width=30mm,minimum height=15mm] (CR) at (5,-3) {\color{black}Fusion Module};
        
        \draw(2,-.5)--(2,4.5)--(11.5,4.5)--(11.5,-.5)--(2,-.5);
        \draw[stealth-stealth,color=black,very thick] (CR)--(5,-0.5);
        \draw[stealth-stealth,color=black,very thick] (CR)--(DM);
        \draw[very thick](Agg)--(event);
        \draw[-stealth,color=black,very thick](event)--(DM);
        \draw[-stealth,color=black,very thick](DM)--(dec);
        \node at (8,0.7) {Multimodal};
        \node at (8,0.4) {Affective };
        \node at (8,0.1) {State Identification};
    \end{tikzpicture}
    }
    \caption{Wearable affect recognition flow: windowing, FBSE–EWT features, per-modality prediction, and uncertainty-weighted decision fusion.}
    \label{fig:pipeline}
\end{figure}
\section{Related Work}
\subsection{Wearable affect detection }
Wearable affect recognition is commonly framed as supervised classification from multimodal physiological Signals, with WESAD dataset becoming a standard benchmark because it provides synchronized chest and wrist modalities and three affective conditions (baseline, stress, amusement) ~\cite{schmidt2018introducing}. Recent work explores both classical Machine learning and deep learning approaches. CNN/LSTM-attention style models are frequently reported as strong baselines for multimodal WESAD classification~\cite{tanwar2024attention}, reflecting the value of temporal modelling when affect clues over time. However, cross-subject generalization remains difficult due to subject-dependent physiology and non-stationary responses. Many methods on WESAD still depends on representation that are either overly aggregated (loosing transient nature) or not designed to be useful across modalities and subjects, leaving space for representation learning that better matches the signal properties of wearable physiology.

\subsection{Frequency and time-frequency representation}

A major line of work relies on frequency domain coefficients, Fourier-domain descriptors and bandpower features are widely used due to simplicity and interpretability. Welch's method is a standard PSD estimator that reduces variance relative to raw FFT periodograms through averaging of modified periodogram~\cite{welch2003use}. These estimators are effective when signals are approximately stationary within short windows, but physiological affect responses can be non-stationary and event-like, where averaging may smooth discriminative structure. Time-frequnecy methods partially address this limitation. STFT provides a localised spectrum but requires a fixed window and thus faces inherent time–frequency resolution trade-offs~\cite{allen2005unified}. Wavelet decompositions (e.g., DWT) provide multi-resolution analysis and are widely used for physiological signals, but still rely on pre-defined basis structure rather than fully data-adaptive segmentation~\cite{mallat1989multifrequency}. Fixed basic spectral and time frequency methods impose analysis choices (band, windows, wavelet families) that can underfit the transient and signal specific structure present in wearable affect data, motivation more adaptive decomposition method that followed the observed spectrum.

\subsection{Adaptive decompositions for non-stationary physiological signals}

To better match non-stationary physiology, research has increasingly used adaptive decompositions such as empirical mode decomposition (EMD/HHT), Varitonal mode decomposition (VMD). EMD decomposes a signal into intrinsic mode functions that capture time varying oscillatory modes  ~\cite{huang1998empirical}. VMD formulates decomposition as a variational optimisation problem with band-limited modes, offering stronger control over mode bandwidth and noise sensitivity \cite{dragomiretskiy2013variational}. EWT learns wavelet bands from the signal’s spectrum via data driven segmentation, providing an adaptive alternative to fixed wavelet banks ~\cite{gilles2013empirical}. These methods motivate our focus to affect related cues in physiology often seen as transient spectral shifts, where adaptive segmentation can improve separability. In short noisy wearable windows, mode separation can still be unstable, and many decompositions remains sensitive to boundary selection and closely spaced components. moreover, their output are not always converted into fixed dimension coefficient in a way that consistently benefits machine learning model across modalities.

\subsection{Multi model fusion}

Multimodal fusion can improve affect recognition by combining complementary physiological cues, as supported by benchmark settings such as WESAD. However, wearable data in practice is affected by artefacts and occasional modality failure, so fusion strategies that operate at the decision level are often preferred over approaches that assume all modalities are always available \cite{gravina2017multi,baltruvsaitis2018multimodal}. In this context, uncertainty aware weighting (e.g., using predictive entropy) provides a principled way to down weight unreliable modality predictions and achieve graceful degradation when signals are missing or noisy ~\cite{aizpurua2018uncertainty}.  



Motivated by these gaps, our approach targets two limitations in wearable affect detection. Representation quality under non-stationary, transient physiological dynamics, and robustness of multimodal integration under missing or unreliable modalities. We introduce an adaptive transient-preserving representation (FBSE–EWT) and couple it with uncertainty and reliability weighted decision-level fusion for multimodal inference in realistic sensing conditions.

\section{Affective State Detection: A three-stage pipeline}

\subsection{Feature Extraction}

\subsubsection{Problem formulation and windowing}

Let subjects be indexed by $s \in \mathcal{S}$ and modalities by $i \in \{1,\dots,M\}$. Each modality provides a discrete time physiological signal
\begin{equation}
x_{s,i}[n], \quad n = 0,\dots, N_{s,i}-1 .
\end{equation}
Signals are segmented into overlapping windows of fixed duration $L$ with overlap ratio $\alpha = 0.75$. With sampling rate $f_i$, the window length is $W = L f_i$ samples and the hop size is $H = (1-\alpha)W$. The $k$-th window is defined as
\begin{equation}
x_{s,i}^{(k)}[n] = x_{s,i}[kH+n], \quad n = 0,\dots, W-1 .
\end{equation}

The WESAD label stream is provided at 700~Hz. Each window receives a label by majority voting with purity threshold $\rho$:
\begin{equation}
y_{s}^{(k)}=\arg\max_{c\in\mathcal{C}} \sum_{t\in \mathcal{I}_k}\mathbf{1}\{\ell_s[t]=c\},
\label{eq:maj_label}
\end{equation}
\begin{equation}
\max_{c\in\mathcal{C}}
\frac{\sum_{t\in \mathcal{I}_k}\mathbf{1}\{\ell_s[t]=c\}}{|\mathcal{I}_k|}
\ge \rho,
\label{eq:purity}
\end{equation}
where $\mathcal{C}=\{\text{Baseline},\text{Stress},\text{Amusement}\}$ and $\mathcal{I}_k$ denotes the corresponding label index range.

\subsubsection{Fourier Bessel representation}

Given a windowed signal $y[n]$ of length $U=W$, the zero-order FBSE represents $y$ using Bessel bases:
\begin{equation}
y[n] = \sum_{m=1}^{U} C_m \, J_0\!\left(\beta_m \frac{n}{U}\right), 
\quad n=0,\dots, U-1 ,
\end{equation}
with coefficients
\begin{equation}
C_m = \frac{2}{U^2 \big(J_1(\beta_m)\big)^2}
\sum_{n=0}^{U-1} n\, y[n] \, J_0\!\left(\beta_m \frac{n}{U}\right),
\end{equation}
where $\beta_m$ is the $m$-th positive root of $J_0(\beta)=0$. Orders $m$ map to physical frequencies via the approximation
\begin{equation}
\beta_m \approx \frac{2\pi f_m U}{f_s},
\qquad \Rightarrow \qquad
m \approx \frac{2 f_m U}{f_s},
\end{equation}
so the FBSE spectrum is $|C_m|$ as a function of $f_m$.

This basis is advantageous for wide-band, non-stationary signals because Bessel bases exhibit AM-like behaviour and can yield compact spectral representations for AFM like components.

EWT constructs empirical wavelet filters whose supports depend on the signal’s spectral content. Boundaries are obtained by locating meaningful minima in a spectrum derived histogram using a scale space persistence criterion, and an automatic threshold (e.g., Otsu) selects which minima are retained. Let the resulting ordered boundaries be
\begin{equation}
0=\omega_0 < \omega_1 < \dots < \omega_N = \pi .
\end{equation}
Given a transition parameter $\xi$, empirical scaling and wavelet functions in the Fourier domain are defined piecewise. These filters form a tight frame under suitable $\xi$, and EWT coefficients are computed by inner products with the empirical wavelets/scaling function, reconstruction follows from the same frame.

In FBSE--EWT, boundary detection and filter-bank construction operate on the FBSE spectrum rather than the conventional FFT spectrum, improving mode separation in challenging cases (e.g., closely spaced components, short-duration activity).

\subsubsection{Feature construction from modes}
\label{subsubsec:features}

Let $K=N$ be the number of resulting modes for window $x_{s,i}^{(k)}$, and let $m_{s,i,k}^{(r)}[n]$ denote the $r$-th reconstructed mode. A fixed-dimensional feature vector $\phi_i\!\left(x_{s,i}^{(k)}\right)$ is formed by aggregating per-mode descriptors. In particular, the mode energy and log-energy are
\begin{equation}
E_r=\frac{1}{W}\sum_{n=0}^{W-1}\big(m_{s,i,k}^{(r)}[n]\big)^2,
\label{eq:mode_energy}
\end{equation}
\begin{equation}
\log\!\left(E_r+\epsilon\right),
\label{eq:mode_log_energy}
\end{equation}
and the mode entropy is computed from the normalised energy distribution
\begin{equation}
p_r[n]=\frac{\big(m_{s,i,k}^{(r)}[n]\big)^2}{\sum_{u=0}^{W-1}\big(m_{s,i,k}^{(r)}[u]\big)^2},
\label{eq:mode_prob}
\end{equation}
as
\begin{equation}
H_r=-\sum_{n=0}^{W-1} p_r[n]\log p_r[n].
\label{eq:mode_entropy}
\end{equation}
This representation preserves transient spectral structure while remaining compatible with standard supervised learning.

\subsection{Single-Sensor Prediction}

The physiological features extracted from each modality are fed into a single-sensor predictor (SSP) to compute the probability of the predicted classes $P_i$ along with the predictor's confidence score $F1_i$. In this study, we adopt a Multi-Layer Perceptron (MLP) as the predictor, denoted as $SSP_{MLP}(\mathbf{S}_i)$, to model non-linear relationships in the feature data collected from a single sensor $\mathbf{S_{i}}$ as presented in Eq.~\ref{eq:MLP}. The choice of MLP is motivated by the nature of the input features, which provide fixed-dimensional representations that already encode transient spectral dynamics. This enables effective learning without the need for explicit temporal models such as CNNs or LSTMs, while maintaining computational efficiency and robustness for relatively small datasets.

In our designed classifier, the MLP consists of multiple fully connected layers with ReLU activation, followed by a softmax output layer that produces class probabilities. Dropout and L2 regularisation are applied to mitigate overfitting. The model is trained using categorical cross-entropy loss and optimised with Adam, with early stopping based on validation performance. We also explore variations in the number of hidden layers and select the configuration that achieves the highest prediction accuracy. The output $P_i$ is interpreted as a probabilistic estimate of the affective state, enabling uncertainty quantification via entropy, while the confidence score $F1_i$ represents the overall reliability of the sensor-specific predictor. These two outputs are subsequently used in the multi-sensor aggregation phase discussed in the following section.

\begin{equation}
    P_i, F1_i \leftarrow SSP_{MLP}(\mathbf{S}_i) \quad \forall\ i\in\{1,2,\dots,n\}
    \label{eq:MLP}
\end{equation}

\subsection{Multi-Sensor Aggregation}
Wearable affect recognition benefits from combining complementary physiological cues, but in realistic settings different modalities can vary in quality across time due to motion artefacts, skin contact variation, or temporary sensor failure. We therefore perform aggregation at the decision level using a weighted average of per-modality probability vectors. The information can be aggregated at the data, feature, or decision level~\cite{gravina2017multi}, and each approach has its challenges and benefits.  In this work, we propose to perform the aggregation on decision-level, and use a weighted average of the individual sensors' decisions.
Let $T\subseteq \{1,2,\dots, n\}$ be a team of sensors, and each sensor feeds data to a single-sensor predictor. The decision $P_i$ yielded by the SSP operating on sensor $i\in T$, contributes to the final aggregated decision according to its {\em entropy} and to the {\em confidence score} of the SSP. Specifically, we obtain the aggregated decision as:
\begin{equation}
    P_T(e) = \frac{1}{\gamma} \sum_{i\in T} \big(1-H(P_i)\big)^{F1_i}\cdot P_i(e) \quad \forall\ e \in \text{classes}
\end{equation}
where $T$ is a team of sensors, $P_i$ is the probability distribution over the classes obtained by sensor $i\in T$; $H(\cdot)$ denotes the information entropy function; $F1_i$ is the F1 score of the single-sensor predictor of sensor $i\in T$; and $\gamma = \sum_{i\in T} H(P_i)^{F1_i}$ is a normalisation factor.
Intuitively, the more certain an SSP is (i.e., the lower the entropy $H(P_i)$), the higher is its contribution to the final decision. Similarly, the more confident the SSP is (i.e., higher confidence score $F1_i$), the more the decision is valued.
Notably, the entropy is predicted-instance specific, while the confidence score is descriptive of the SSP in general. 
As such, using the entropy as the base and the confidence score as the power, we regulate the influence of each individual decision, prioritising primarily according to the entropy, and secondly according to the confidence score.

\begin{figure}[]
    \centering
    \hspace{-1cm}
    \begin{tikzpicture}
    \pie[radius=2.5,color={green!60!white, yellow!60!white, red!40!white}]{48.39/D>F, 35.48/D=F, 16.13/D<F}
    \end{tikzpicture}
    \caption{Overall Accuracy}
    \label{fig:accuracyPieChart}
\end{figure}
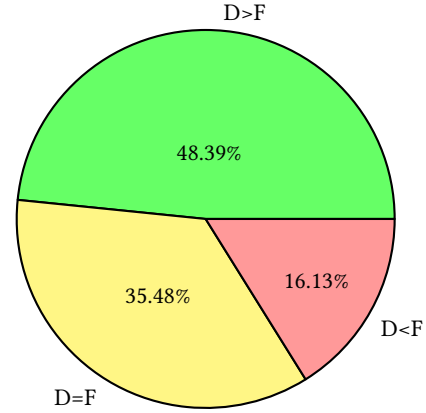

\begin{figure}
    \centering
       \scalebox{0.9}{
    \begin{tikzpicture}
    \begin{axis}[
        ybar,
        enlargelimits=0.15,
        legend style={at={(0.5,-0.15)},
          anchor=north,legend columns=-1},
        ylabel={accuracy},
        symbolic x coords={all, quartet, triad, pair, single},
        xtick=data,
        ymin=0.5,
        ]
    \addplot+[error bars/.cd, y dir=both,y explicit] coordinates {
        (all,1) 
        (quartet,0.96) +- (0.0,0.059628)
        (triad,0.88) +- (0.0,0.132591)
        (pair,0.866667) +- (0.0,0.143329)
        (single,0.706667) +- (0.0,0.238514)
     };
      \addplot+[error bars/.cd, y dir=both,y explicit] coordinates {
        (all,0.933333) 
        (quartet,0.906667) +- (0.0,0.036515)
        (triad,0.886667) +- (0.0,0.070623)
        (pair,0.807407) +- (0.0,0.147573)
        (single,0.706667) +- (0.0,0.238514)
     };
    \end{axis}
    \end{tikzpicture}
    }
\caption{Team level accuracy}
\label{fig:results:team} 
\end{figure}
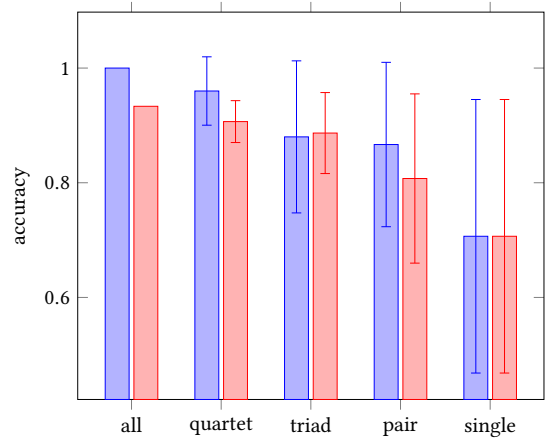

\section{Implementation and Results}
We performed some preliminary evaluation using the WESAD dataset~\cite{schmidt2018introducing} for three-class affect recognition (baseline, stress, amusement). We use five sensors, electrocardiogram (ECG), electrodermal activity (EDA), blood volume pulse (BVP), electromyogram (EMG) and three-axis acceleration, for 15 subjects. In these experiments, we compare the decision-level aggregation against the feature-level one. Specifically,
Figure~\ref{fig:accuracyPieChart} shows the overall accuracy in correctly predicting an class when the aggregation was performed at the feature (F) and decision (D) levels. We note that in $\approx84\%$ cases, decision aggregation is as good as or better than feature aggregation. 
Next, we explore the impact of the {\em team size} on the accuracy. That is, instead of considering all five sensors, we test the accuracy of a team consisting of varying team-size combinations of sensors (including teams of a single sensor as well).
Figure~\ref{fig:results:team} shows the comparison of feature-level and decision-level aggregation of all the team sizes.
Notably, regardless of the team size, decision-level aggregation is better. Additionally, we observe that when more sensors participate in the aggregation results in better accuracy. 

To conclude, our preliminary results indicate that the proposed decision-level aggregation achieves better accuracy than the typical feature-level one.
Decision-level aggregation is more generic and can be used to aggregate a diverse set of sensors without expert knowledge regarding specific signal features. It is also robust, since if one of the sensors fail, the decisions of the other sensors can still be aggregated.

\section{Conclusions and Future Work}
This paper presented a three-stage pipeline for wearable affect recognition that combines a non-stationary representation (FBSE – EWT) with uncertainty and reliability weighted decision-level fusion. The proposed approach is motivated by deployment realities in societally important health and care settings, where physiological sensing is noisy, non-stationary, and occasionally incomplete, and where downstream systems should act on calibrated beliefs rather than overconfident predictions. Preliminary experiments on WESAD (baseline, stress, amusement) indicate that decision-level fusion is competitive with, and often superior to, feature-level fusion across sensor-team sizes, suggesting improved robustness under heterogeneous sensing conditions.

In future, we plan to work and extend this line of work in some of the following ways: {\em(i)} use multi-modal input, like physiological sensors, video, audio, images and natural language input, {\em(ii)} augment the data by using data generators for different sensors, {\em(iii)} detect a variety of relevant classes, {\em(iv)} deploy the model in different types of care homes. {\em(v)} process temporal information, {\em(vi)} communication and coordination among diverse robots from different manufacturers and having a different set of sensors. We also plan to validate on additional datasets and integrate the resulting uncertainty-aware affect estimates into agent-based decision support pipelines for stress-aware interventions and continuous wellbeing monitoring.

\begin{acks}
This work is supported by Responsible Ai UK (\href{https://rai.ac.uk/}{RAi UK}) (EP/Y009800/1) coordinating keystone project on ``Embodied AI in Social Spaces: Responsible and Adaptive Robots in Complex Settings''.
\end{acks}

\balance






\bibliographystyle{ACM-Reference-Format} 
\bibliography{sample}


\end{document}